\documentclass[aps, prd, amsmath, floats,floatfix, twocolumn,superscriptaddress, nofootinbib, showpacs]{revtex4}
\usepackage{amsmath} % Useful for equation alignment
\usepackage{longtable}
\usepackage{verbatim} % Allows to have comments over multiple lines
\usepackage{mathrsfs}
\usepackage{amsfonts}
\usepackage{dsfont}
\usepackage{color}
\usepackage{graphicx}

\newcommand{\be}{\begin{equation}}
\newcommand{\ee}{\end{equation}}
\newcommand{\bea}{\begin{eqnarray}}
\newcommand{\eea}{\end{eqnarray}}

\begin{document}

\title{Can the dark matter halo be a collisionless ensemble of axion stars?}

\author{J. Barranco}
\email{jbarranc@fisica.ugto.mx}
\affiliation{Division de Ciencias e Ingenier\'ias,  Universidad de Guanajuato, Campus Leon,
 C.P. 37150, Le\'on, Guanajuato, M\'exico.}

\author{A. Carrillo Monteverde}
\email{alcarrillo@fisica.ugto.mx}
\affiliation{Division de Ciencias e Ingenier\'ias,  Universidad de Guanajuato, Campus Leon,
 C.P. 37150, Le\'on, Guanajuato, M\'exico.}

\author{D. Delepine}
\email{delepine@fisica.ugto.mx}
\affiliation{Division de Ciencias e Ingenier\'ias,  Universidad de Guanajuato, Campus Leon,
 C.P. 37150, Le\'on, Guanajuato, M\'exico.}

\date{\today}

\begin{abstract}
If dark matter is mainly composed of axions, the density distribution can be nonuniformly distributed, 
being clumpy instead. By solving the Einstein-Klein-Gordon system of a scalar field
with the potential energy density of an axionlike particle, we obtain the maximum
mass of the self-gravitating system made of axions, called axion stars.
The collision of axion stars with neutron stars may release the energy of axions
due to the conversion of axions into photons in the presence of the neutron star's magnetic field.
We estimate the energy release and show that it should be much less than previous estimates \cite{Iwazaki:1999ub,Iwazaki:1999my}.
Future data from femtolensing should strongly constrain this scenario.
\end{abstract}
\maketitle
\section{Introduction}

Nowadays one of the most interesting problems in physics is to reveal the nature of dark matter (DM).
The existence of DM is supported by large amount of astrophysical observations, on both galactic and cosmological
scales \cite{Bertone:2004pz,Komatsu:2010fb}. Since those effects are purely gravitational,
the standard candidate is a massive particle with null or very weak interaction with the
rest of the particles of the standard model of particles.
Two of the leading candidates for dark matter are axions and weakly-interacting massive particles such as
neutralinos \cite{Bertone:2010zz}.

The axion is the pseudo Nambu-Goldstone boson generated in the spontaneous breaking of the
$U(1)$ Peccei-Quinn (PQ) global symmetry \cite{Weinberg:1977ma,Wilczek:1977pj,Peccei:1977hh}.
PQ symmetry was introduced to explain the smallness of
the strong CP violation in QCD. Initially, this new field was assumed to be "invisible" as very weakly
interacting  but in 1983, it has been shown that axion couples to two photons through Primakoff
effects \cite{Primakoff:1951pj} and could be detected through its conversion into a photon in presence of a
strong magnetic field \cite{Sikivie:1983ip}.

Most of the direct detection methods for the axion are based on the
conversion of axions into photons \cite{Smith:1988kw,Avignone:1986vm, Arisaka:2012pb}. But it is also possible to constrain the axion mass and coupling through astrophysics observation. As stars usually have strong magnetic fields, the Sun or any other astrophysical object as dwarf star, supernova, etc. 
could produced axions in its core during cooling processes and transform them into photons through Primakoff effects.  The  search for  photons 
produced in stars through axion conversion can be used to put an upper limit on axion mass  $m_a$ and coupling $f_a$, where $f_a$ defines the scale of $U(1)_{PQ}$ symmetry breaking: $m_a \leq 10^{-3} \mbox{eV}$  and $f_a \geq 10^{9} \mbox{GeV}$ \cite{Iwamoto:1984ir,Raffelt:1987yt,Brinkmann:1988vi,Burrows:1988ah,Raffelt:1993ix,Janka:1995ir,Umeda:1997da,Hanhart:2000ae,Raffelt:2011ft,FerrerRibas:2012ru}. Cosmological observations impose constraints on $m_a$ to be bigger than $10^{-6}\mbox{eV}$ and $f_a \leq 10^{12} \mbox{GeV}$ \cite{Preskill:1982cy,Abbott:1982af,Turner:1989vc,Kim:2008hd}. Therefore, cosmological and astrophysical observations constrain the range of $m_a$ to be
\begin{equation}
10^{-6} \mbox{eV} < m_a< 10^{-3} \mbox{eV}\,.
\end{equation}
So, an axion mass within this range could be a nice candidate for  cold dark matter \cite{Sikivie:2006ni,Hwang:2009js} as it could form Bose-Einstein condensate \cite{Sikivie:2009qn}.

The rate of detection needs a prescription on the dark matter distribution in the galactic halo,
and in particular, the local dark matter density.
There are arguments in favor of a nonuniform distribution of dark matter in the galaxy \cite{Dalal:2001fq,DeLucia:2003xe,Diemand:2008in,Keeton:2009ua,Berezinsky:2003vn,Berezinsky:2005py}.
For instance, early fluctuations in the dark matter may go
nonlinear long before photon decoupling, producing density-enhanced dark matter clumps.
The evolution of the axion field at the QCD transition epoch
may produce gravitationally bound miniclusters of axions \cite{Hogan:1988mp,Schmid:1998mx}.
Such minicluster, due to collisional
$2a \to 2a$  processes, may relax to a selfgraviting system \cite{Kolb:1993zz,Kolb:1993hw,Tkachev:1991ka}.
These self-gravitating systems made of axions are  simply  called axion stars (AS).
Previously, some authors explored the possibility that such axion stars may have some
observable consequences. For instance, the collisions between axions stars and neutron stars
may induce gamma ray burst \cite{Iwazaki:1999vi}, ultra-high energy cosmic rays \cite{Iwazaki:2000hq},
or emission of x rays by neutron stars \cite{Iwazaki:1999ub}.
All of these observables have in common the possibility that the axion may oscillate into a photon.

In this paper, we study the self-gravitating system
made of axions with a realistic potential for the scalar field as in \cite{Barranco:2010ib}  contrary to \cite{Iwazaki:1999vi,Iwazaki:2000hq,Iwazaki:1999ub}. In section II, we obtain the typical Mass-density curve varying the central density. This curve exhibits a maximum mass  which depends on the value of the axion mass and the central density. Section III is dedicated to update signatures of the AS in the Galactic Halo due to AS collisions with neutron stars (NS). The huge magnetic field in NS, $B \sim 10^{8}$ Gauss, may convert to photons all axions involved in the collisions.  In this section, an estimate of the total radiated energy is computed. Finally, we present our results and compared to previous estimates in
Section IV.

\section{Axion Stars}
\subsection{On the formation of axion stars}
The formation of axion miniclusters was studied in \cite{Hogan:1988mp,Kolb:1993zz,Kolb:1993hw}.
They did a numerical study of the evolution of a spherically symmetric axion fluctuation
in an expanding universe at the QCD epoch, where the mass of the axion effectively switches on.
The axion potential considered was
\begin{equation}\label{potential}
V(\phi)=m_a^2 f_a^2 \left[ 1 - \mbox{cos}\left( \frac{\phi}{f_a} \right)\right] \,,
\end{equation}
where $\phi$ is the axion field, $f_a$ the energy scale of decoupling, and $m_a$ the axion mass.
The study was done for a wide range of initial conditions and a common feature
of the final density distribution is that it developed a sharp peak in the center, i.e.
due to the attractive self-interaction, some non-linear effects induce
axion perturbations that lead to very dense axion miniclusters.
If such density peaks are high enough, gravitational collapse and subsequent virialization
can occur. This relaxation is due to $2a \to 2a$ scattering and the associated
relaxation time is smaller than the age of the universe if the energy density of the minicluster
is $\rho > 10^{10}{\rm eV}^4$ \cite{Kolb:1993zz}.
As we will see later, the axion star fulfills the conditions needed, such as
relaxation, and annihilation processes prevent a gravothermal catastrophe for the axion
minicluster \cite{Tkachev:1991ka}.

An improved numerical resolution of the axion mini-clusters revealed that around $\sim 10\%$ of
the axion density will be in miniclusters with densities that can virialize to coherent axion fields
on a self-gravitational well, i.e. axions stars \cite{Kolb:1995bu}.

Other mechanisms that may lead the formation of axion miniclusters and dark matter miniclusters
were proposed in \cite{Schmid:1998mx}. There, at the time where the phase transition from a quark-gluon plasma
to a hadron gas take place, the spectrum of density perturbations
develops peaks and dips produced by the growth of hadronic bubbles.
Afterwards, kinematically decoupled cold dark matter falls into
the gravitational potential wells provided from those peaks, leading the formation of
dark matter clumps with masses $<10^{-10}M_\odot$.

\subsection{Properties of the axion stars}

\begin{figure}
\includegraphics[angle=0,width=0.48\textwidth]{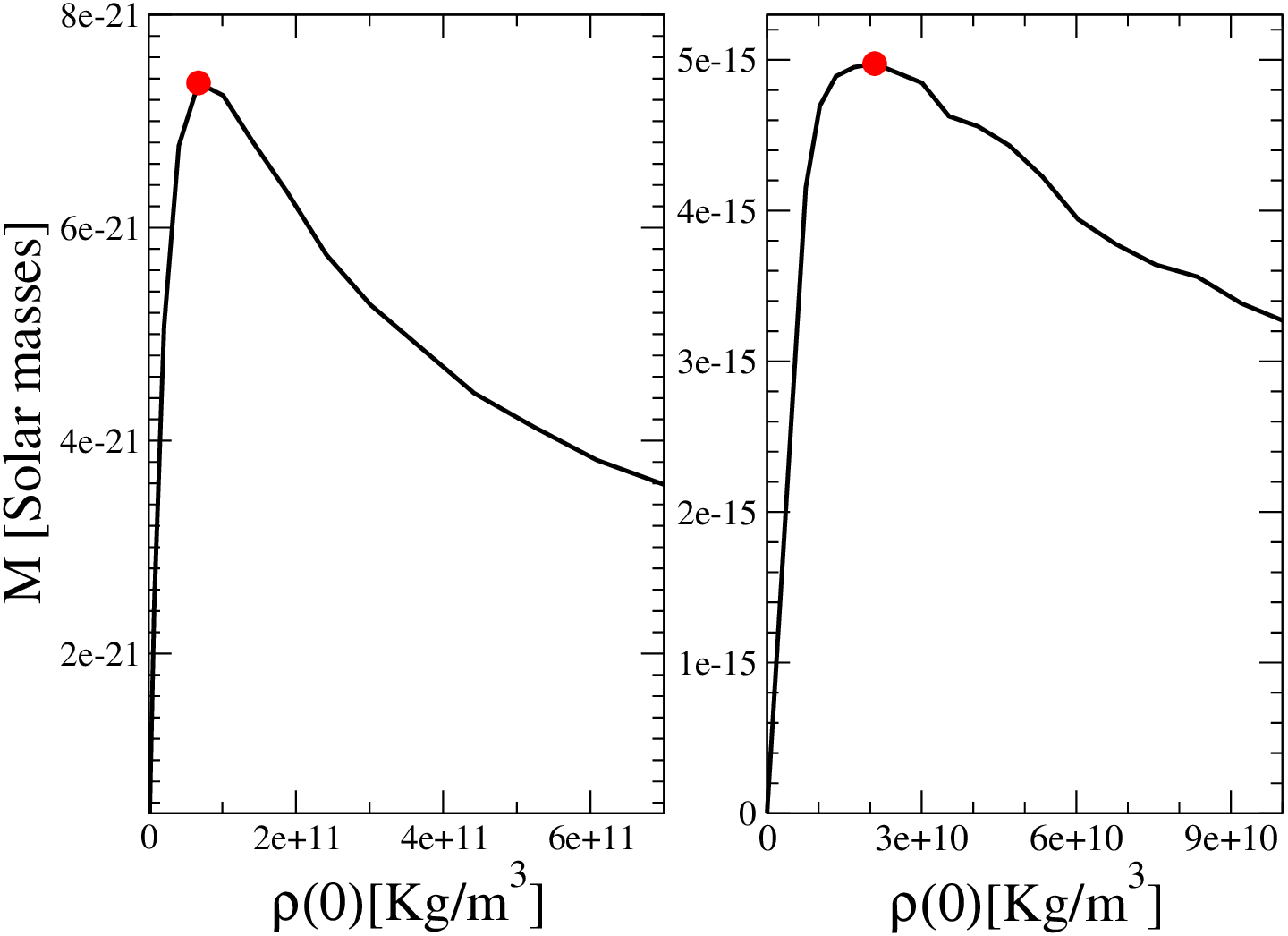}
\caption{Mass of axion star. Left Panel: $m_a=10^{-3}$eV. Right Panel: $m_a=10^{-5}$eV.}\label{Fig1}
\end{figure}
\begin{figure}
\includegraphics[angle=0,width=0.48\textwidth]{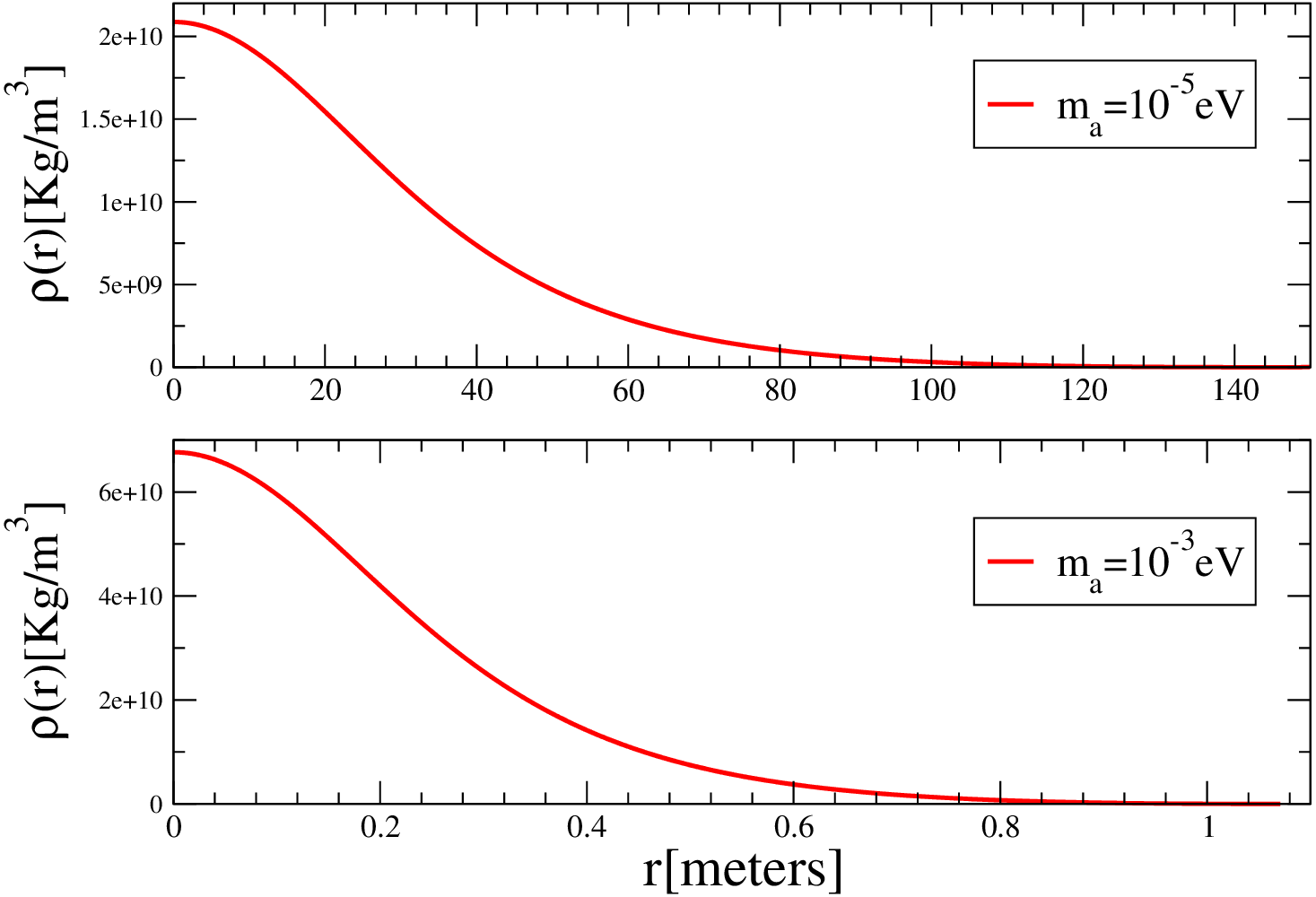}
\caption{Density profile. Top Panel: $m_a=10^{-5}$eV. Bottom Panel: $m_a=10^{-5}$eV.}\label{Fig2}
\end{figure}

The previous picture of the formation of axion stars did not study the final
self-gravitating system made of coherent axion fields. In order to do so,
we have solved the Einstein-Klein-Gordon equation in the semiclassical limit:
\begin{equation}\label{EKG}
G_{\mu\nu}=8 \pi G \langle \hat T_{\mu\nu} \rangle\,,
\end{equation}
\begin{equation}
\left(\Box-\frac{dV(\Phi)}{d\Phi^2}\right)\Phi=0\,,
\end{equation}
where the source of Einstein equations, the energy momentum tensor
$\langle \hat T_{\mu\nu} \rangle$, is the
average over the ground state of a real, quantized  scalar field $\Phi(r,t)$
with potential given by the energy potential in Eq. (\ref{potential}).
We  work in a spherically symmetric metric
$ds^2=B(r)dt^2-A(r)dr^2-r^2(\sin^2\theta d\phi^2+d\theta)\,$,
and to assume a harmonic dependence of the field $\Phi(r,t)=e^{i\omega t}\phi(r)$.
The EKG system reduces to the following system of equations:
\begin{eqnarray}
a'+\frac{(1-a)a}{x}+(1-a)^2x {} \nonumber \\  {}\times
\Big[\left(\frac{1}{\tilde B}+1\right)m_a^2\sigma^2-
\frac{m_a\sigma^4}{12}+\frac{\alpha \sigma'^2}{1-a}+\frac{\sigma^6}{360}\Big]&=&0\nonumber\\
\tilde B+\frac{a\tilde B}{x}-(1-a)\tilde Bx {} \nonumber \\  {} \times
\Big[\left(\frac{1}{\tilde B}-1\right)m_a^2\sigma^2+
\frac{m_a\sigma^4}{12}+\frac{\alpha \sigma'^2}{1-a}-\frac{\sigma^6}{360}\Big]&=&0\nonumber\\
\sigma''+\left(\frac{2}{x}+\frac{\tilde B'}{2\tilde B}-\frac{a'}{2(1-a)}\right)\sigma' + {}
\nonumber \\  {} + \frac{1-a}{\alpha}\Big[\left(\frac{1}{\tilde B}-1\right)m_a^2\sigma^2
+\frac{m_a\sigma^3}{6}-\frac{\sigma^5}{120}\Big]&=&0\,,\label{sistema}
\end{eqnarray}
where we have defined $\phi=\frac{f_a}{\sqrt{m_a}}\sigma$, $\tilde B=\frac{m_a^2 B}{\omega^2}$,
$r=\frac{m_p}{f_a}\sqrt{\frac{m_a}{4 \pi}}x$, $a=1-A$ and $\alpha=\frac{4\pi}{m_a}\frac{f_a^2}{m_p^2}$,
as  was done in \cite{Barranco:2010ib}. In the present work, we extend the results of \cite{Barranco:2010ib}
by solving the same system but for a wider range of masses of the axion field and many initial values of the
axion field at the center $\sigma(x=0)$.
Since the axion mass is restricted to be $10^{-6}\mbox{eV}<m_a<10^{-3}\mbox{eV}$, we
show in Fig. \ref{Fig1} the resulting masses of the AS as a function of the value of the central
density $\rho(0)=m_a^2 \sigma(0)^2/2$.
The mass-density plot shows that there exists a maximum mass for a particular value of the central
density. The density profile of those maximum mass configurations are shown in Fig. \ref{Fig2} and
the numerical values are reported in Table \ref{physicalvalues}.
Those configurations are our benching mark in order to study possible astrophysical signatures.
 Once we have computed the maximum masses of AS for different choices of $m_a$, we can come back to
the question of their plausibility. In \cite{Tkachev:1991ka}, it was found that AS formation is possible if the
mass in the core
region $M$, the escape velocity $v_e$ and the axion self-coupling $f_a$ satisfy the relation
$M<10^8 v_e^{5/2}(f_a/10^{10}\mbox{GeV})^{-1/4}$.
By using the lower value of $f_a=10^9$ GeV \cite{Iwamoto:1984ir,Raffelt:1987yt,Brinkmann:1988vi,Burrows:1988ah,Raffelt:1993ix,Janka:1995ir,Umeda:1997da,Hanhart:2000ae,Raffelt:2011ft,FerrerRibas:2012ru}, and the
 maximum mass we have obtained for an AS
for $m_a=10^{-5}$ eV, and $v_e=\sqrt{2 GM/r}$, we can check that such condition is fulfilled,
and then, that the relaxation time is lower than the age of the universe.
For the annihilation processes, one may be worried for the
stimulate decays into photons. It occurs when the condition $\Gamma_\pi m_p^2 v_e f_\pi /(R m_\pi^4 f_a) > 1$
is fulfilled \cite{Tkachev:1987cd,Seidel:1993zk}, but such a condition implies densities $\rho > 10^{15}$ Kg/ m$^3$ for
$m_a=10^{-5}~$eV. As we can see from Fig. \ref{Fig2} the AS's density  is always below $\rho(0)<2.1 \times 10^{10}
\left({\rm Kg}/\mbox{m}^3\right)$. Hence, no stimulated decays of axions is produced.

\begin{table}
\caption{Masses and typical radio $R_{99}$ for the maximum mass configurations shown in Fig. \ref{Fig1} where $R_{99}$ is
the radius of the AS where $99\%$ of the mass is contained}\label{physicalvalues}
        \begin{tabular}{c|c|c|c}
            Axion mass (eV)&$\rho(0) \left({\rm Kg}/\mbox{m}^3\right)$ & Mass ($M_\odot$) & $R_{99}$ (meters) \\
            \hline
            $m_a=10^{-5}$&$2.1 \times 10^{10}$  & $5.0\times 10^{-15}$ & $119.40$ \\
            $m_a=10^{-3}$&$6.8 \times 10^{10}$ & $7.4\times 10^{-21}$ & $0.89$ \\
            \hline
         \end{tabular}
\end{table}

\section{Possible axion star signatures in the galactic halo}
The dark matter halo of galaxies can be considered as a collisionless ensemble of mini-MACHOs, as far as this
hypothesis is not in contradiction with the limits imposed by microlensing or gravothermal instability.
In particular, scalar field mini-MACHOS has been considered as a possible realization if
their masses are below $10^{-7}M_\odot$ \cite{Hernandez:2004bm} to evade the microlensing limits.Recently, it was shown that mini-MACHOS with masses below $10^{-10}M_\odot$ have not measurable dynamical consequences in internal
Solar System dynamics (planets, Earth-Moon) \cite{GonzalezMorales:2012ab}.
In turns out from our analysis on the maximum mass for an axion star, for $m_a=10^{-5}~$eV,
axions stars may play the role of such mini-MACHOs.
In addition to microlensing, femtolensing could be a useful tool in order to look for
axion stars \cite{Kolb:1995bu}. Recently, new limits from femtolensing of gamma ray burst
provides new evidence that primordial black holes in the mass range $10^{-16}-10^{-13}M_\odot$ do not
constitute a major fraction of the dark matter \cite{Barnacka:2012bm} and once again, the maximum mass
of the axion star are consistent with those limits.
In the case for an axion mass of $m_a=10^{-3}$eV, the maximum AS mass is far to be excluded from
these current limits from femtolensing of GRBs since $M_{max} \sim 10^{-21}M_\odot$ but it is encouraging
that in the future, femtolensing measurement may confirm or deny the possibility that the dark matter halo
may be made of an ensemble of axion stars.
In addition to AS, other types of dark matter clumps may play the role of mini-MACHOS. In particular
if dark matter is a non self-annihilating fermion, for some values of its mass it may fulfill the requirements to be
a mini-MACHO \cite{Narain:2006kx}.
For a self-annihilating fermion as the neutralino, neutralino stars \cite{Berezinsky:1996eg,Bergstrom:1998jj,Ren:2006tr},
have been proposed. Actually, the maximum mass for a neutralino star is $\sim 10^{-7}M_\odot$
making them suitable dark matter mini-MACHOS candidates, although their stability
is questionable \cite{Dai:2009ik}.

The clumping of dark matter may change the expected number of events for direct dark matter searches,
since the local density may be largely affected at the vicinity of the Earth.
Nevertheless, the clumpy dark matter may offer new ways for indirect detection.
In the case of the axion, early studies of luminous axion clusters where done in the 1980s \cite{Kephart:1986vc}.
In this work we  reconsider the proposal of Iwazaki \cite{Iwazaki:1999vi} where axion stars interact with
the strong magnetic field of a neutron star producing photons, now we consider the
mass and radius of the AS obtained as general relativity indicates by solving Eq. \ref{sistema}.
\begin{figure}
\begin{center}
\includegraphics[angle=0, width=.49\textwidth]{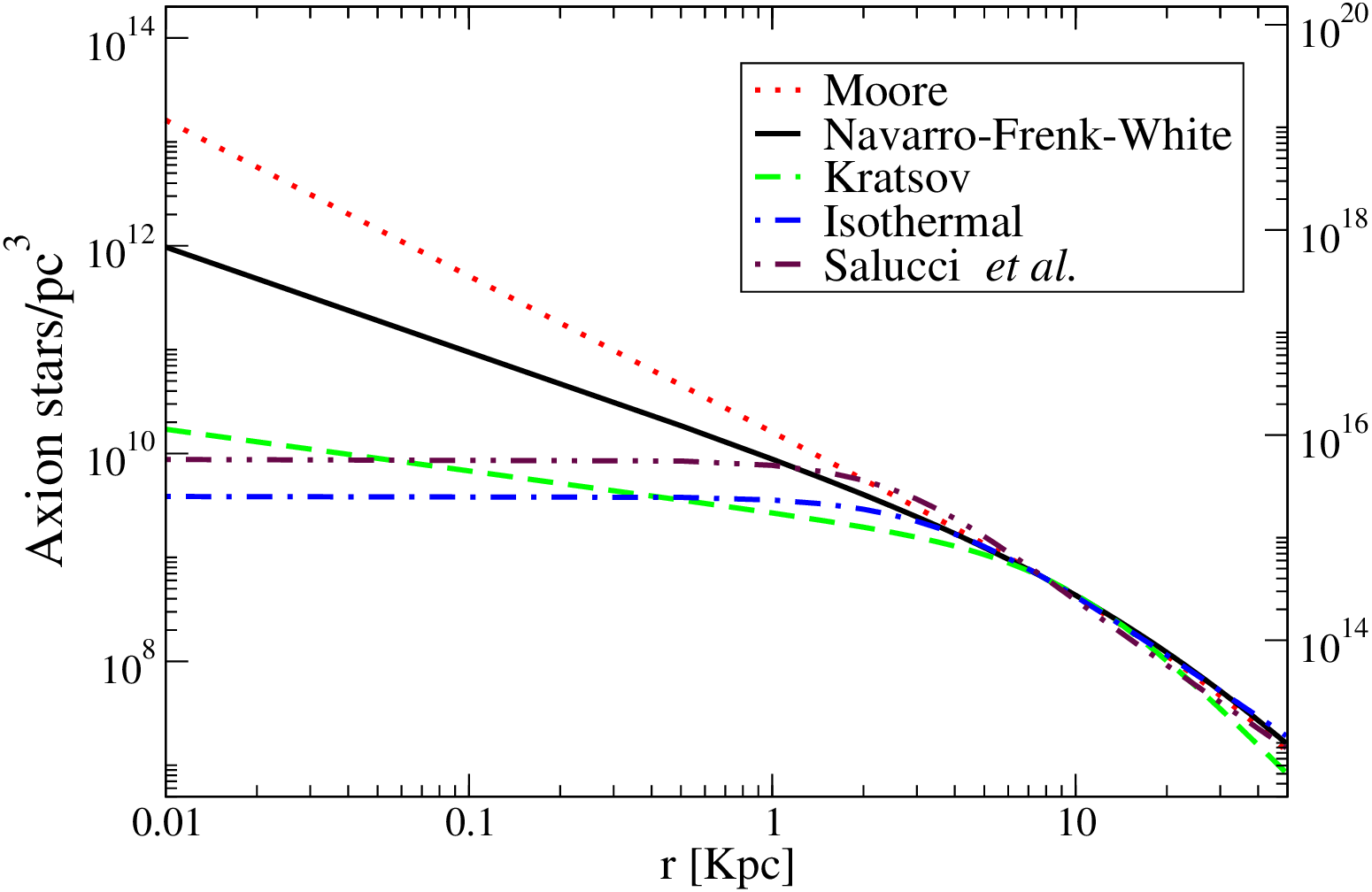}
\caption{Number of axion stars. Left y-axis corresponds to the case for
a mass of $10^{-15}M_\odot$, while right y-axis corresponds to AS mass of $10^{-21}M_\odot$.}\label{ASnumber}
\end{center}
\end{figure}
\begin{table}
\caption{Energy radiated for an AS with the maximum mass configurations shown in Fig. \ref{Fig2}
for a NS magnetic field $B=10^{8}$G and an electric conductivity $\sigma=10^{26}s^{-1}$. \cite{Baiko:1995qg}.}
\label{Table2}
        \begin{tabular}{c|c|c|c}
            Axion mass (eV)&$f_a$ (GeV)& Mass ($M_\odot$) & W ($\mbox{erg/s}$) \\
            \hline
            $m_a=10^{-5}$& $ 6 \times 10^{9}$  & $5.0\times 10^{-15}$ & $5.3\times 10^{43}$ \\
            $m_a=10^{-3}$& $ 6 \times 10^{11}$ & $7.4\times 10^{-21}$ & $7.9\times 10^{37}$ \\
            \hline
         \end{tabular}
\end{table}
First, we shall estimate the number of AS in a galactic halo using dark matter density profiles.
The AS number density is modified by the tidal destruction of
axion miniclusters which occurs within a hierarchical model of clump structure. The
tidal destruction arise by the gravitational interaction of two clumps
passing near each other or when the minicluster falls into the gravitational field of the host.
The clump survival  probability, as a function of the mass and radius of the clump,
for non dissipative DM particles, was computed in
\cite{Berezinsky:2003vn,Berezinsky:2005py}. It was found that only $0.1-0.5\%$ of the small
clumps survive the stage of tidal destruction.
In Fig. \ref{ASnumber} it is shown  in the left y axis (right y axis)
the number of axion star with a mass of $10^{-15} M_\odot$ ($10^{-21}M_\odot$) for five different DM density profiles:
the Navarro-Frenk-White profile \cite{Navarro:1995iw}, the Moore profile \cite{Moore:1999gc}, the Kravtsov profile 
\cite{Kravtsov:1997dp},
the modified isothermal profile \cite{Bergstrom:1997fj} and the Salucci {\it et al.} profile
\cite{Salucci:2007tm,Donato:2009ab}. The values reported in Fig. \ref{ASnumber} included a factor $10^{-3}$
to account the tidal destruction of clumps and a factor $10^{-1}$ due to the fact that around $\sim 10\%$ of
the axion density will be in miniclusters with densities that can virialize to coherent axion fields \cite{Kolb:1995bu}.
Notice the high number of axion stars at the center of the galaxy.
Furthermore, it is expected that $\sim .1\%$ of a galaxy's total stars will
finish their life as neutron stars. The Milky-Way, for instance, could have around
$\sim 10^{9}$ neutron stars. The high number of axion stars (see Fig. \ref{ASnumber}) and the
expected number of neutron stars implies a non-zero probability that AS  collide with a neutron star
and dissipate their energy under the effect of the neutron stars' magnetic field.
In order to estimate the amount of energy radiated
by this process, we  closely follow  Ref. \cite{Iwazaki:1999vi}.
Starting with the Lagrangian
\begin{equation}\label{axion-photon-lagrangian}
\mathcal{L}=\frac{1}{2}(\partial^\mu a \partial_\mu a-m^2 a^2)
-\frac{1}{4}\frac{ c a}{f_a}F_{\mu \nu}\tilde F^{\mu \nu}-
\frac{1}{4}F_{\mu \nu}F^{\mu \nu}\,,
\end{equation}
here $c$ is a coupling constant of order unity, $a$ is the axion field
and $F^{\mu\nu}$ is the electromagnetic stress tensor and $\tilde F^{\mu\nu}$
its dual \cite{Sikivie:1983ip,Raffelt:1987im}.
And expressing this Lagrangian  (\ref{axion-photon-lagrangian})
in terms of the electric and magnetic fields $\vec E, \vec B$.
\begin{equation}\label{newlagrangian}
\mathcal{L}_{a\gamma\gamma}=\frac{c \alpha}{f_{a}} a \vec E \cdot \vec B\,
\end{equation}
we can now derive a ``modified'' Gauss law, namely:
\begin{equation}
\partial \vec E= \frac{-c \alpha}{f_{a}} \vec \partial \cdot (a \vec B)\,.
\end{equation}
Thus, the axion field $a$ may induce an electric field $\vec E_a=-c \alpha a \vec B/f_a$.
If the neutron star has an electric conductivity $\sigma$, an electric current
$J_m=\sigma E_a$ is produced.
The dissipation in the magnetized conducting media, with average $\sigma$
electric conductivity, is evaluated using Ohm's law $W=\int_{AS}\sigma E_a^2d^3x$,
which, with help of the modified Gauss Law, can be directly related with the axion field
and the neutron star magnetic field:
\begin{equation}\label{dissipated_energy}
W=4 \pi \frac{c^2 \alpha^2 B^2 \sigma}{f_a^2} \int_0^R a(r)^2 r^2 dr\,.
\end{equation}

\begin{figure}
\includegraphics[angle=0, width=.49\textwidth]{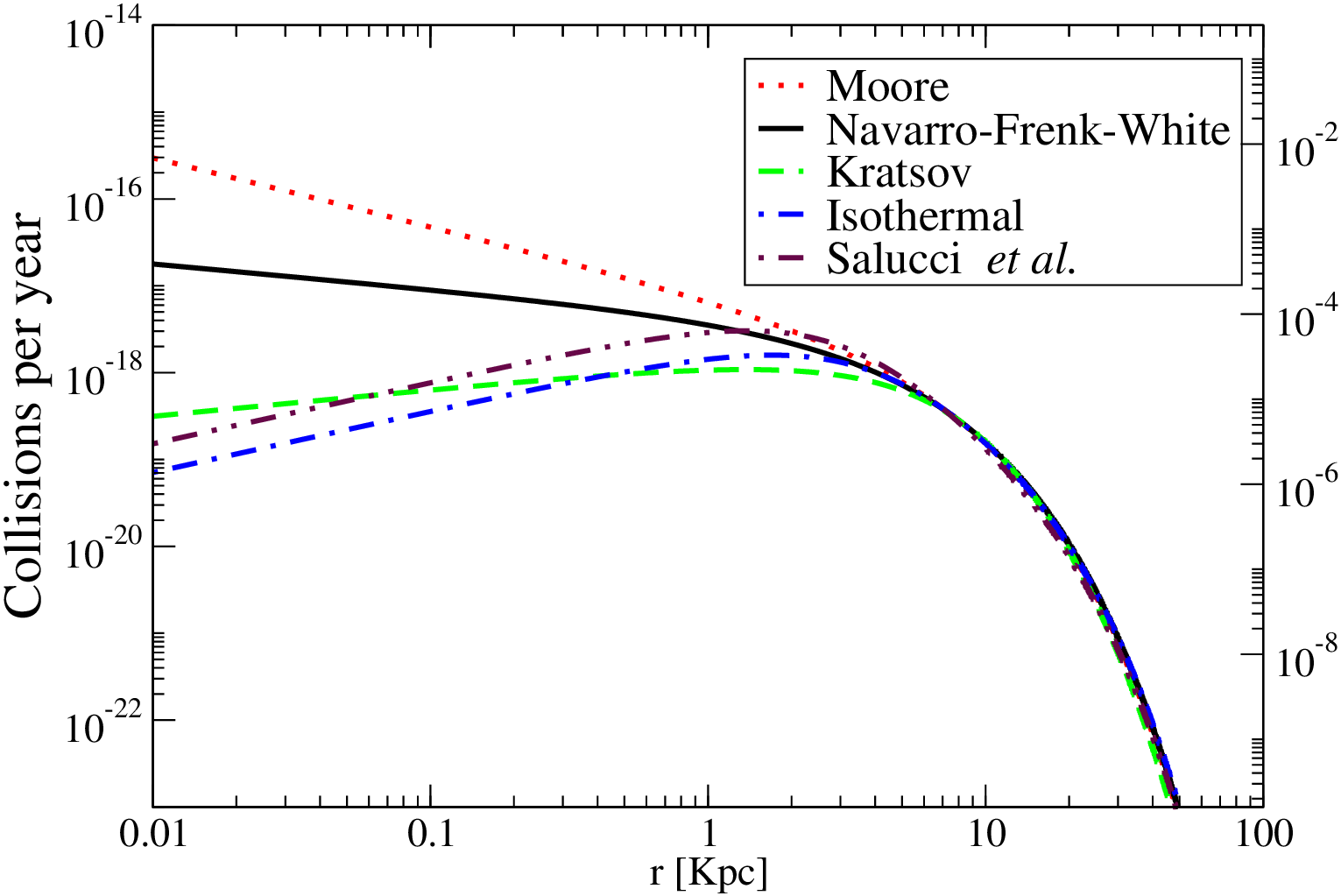}
\caption{Number of collisions of axion stars with neutron stars per year
as a function of radial distance to the center. Left y-axis
corresponds to an axion star mass $5.\times 10^{-15}M_\odot$ and an
effective cross section $S=\pi R_{99}^2$. Right y-axis
corresponds to an axion star mass $7.4\times 10^{-21}M_\odot$ and
$S=\pi L_c^2$, $L_c$ defined in the text.}\label{energy}
\end{figure}

Taking the two profiles for the axion field shown in Fig. \ref{Fig2},
we  perform the integration of Eq. (\ref{dissipated_energy}) numerically.
The radiated energy is shown in Table \ref{Table2}.
Finally, we  estimate the rate of collisions, such as we
estimate the luminosity radiated every second.
The number of collisions per $pc^{3}$ per second will be
\begin{equation}
R_c=n_{AS}(r)\times \rho_{NS}(r)\times S \times v \,,
\end{equation}
where $n_{AS}$ is the number of AS per $pc^3$ as a function of the distance to the
galactic center (Fig. \ref{Fig1}),
$\rho_{NS}$ is the probability  to find a Neutron Star at that point,
and we will assume this distribution to be \cite{Taani:2012xp}
\begin{equation}
\rho_{NS}(r)=A r^{\alpha-1}/\lambda^\alpha e^{-r/\lambda}\,,
\end{equation}
where $\alpha \simeq 1.7$ and $\lambda (\mbox{kpc})= 5.21$, as given in Ref. \cite{Taani:2012xp}.
For simplicity, the $z-$dependence  is neglected. This assumption implies that
NS distribution is assumed to be spherically symmetric.
$A$ is  the normalization constant defined such that
\begin{equation}
10^{9}\mbox{NS}=\int A \rho_{NS}(r)
r^2\sin^2\theta dr d\theta d\phi\,.
\end{equation}
The velocity $v$ of AS is assumed to be around  $v=2\times 10^{7}$cm/s.
Finally, $S$ is the cross section which can
be computed either as the effective area of the AS, i.e. $S_{99}=\pi R_{99}^2$, or we can define
a distance $L_c$ such that the kinetic energy of the axion star is equal to the
potential energy between the AS and the neutron star, $GM_{AS}M_{NS}/L_c$. When such
condition happens, we can say the collision occurs. Thus $S_{L_c}=\pi L_c^2$.
The number of collision per year $R_c$ is shown in Fig. \ref{energy} for two cases:
The left y-axis illustrate the case for the maximum mass of the axion star $M_{AS}=10^{-15}M_\odot$
and $S=S_{99}$ while the left y-axis shows the case where $M_{AS}=7.4\times 10^{-21}M_\odot$ and effective
cross section $S=S_{L_c}$. $R_c$ is very small in the first case, while the second case is more
optimistic, $\sim$ one per century per galaxy. Nevertheless, even in this ideal case,
the effects will be hard to detect. The release of energy due to the collision is shown for both maximum cases
in Table \ref{Table2}. We can estimate the change in temperature if all energy
is transferred to the NS. By assuming that the thermal energy in the NS is
$U=6 \times 10^{47} \mbox{erg}\left(M_{NS}/M \right)\left(\rho/\rho_n\right)^{-2/3}\left(T/10^{9}\mbox{K}\right)^2$
\cite{Iwazaki:1999ub} with $\rho_n=2.8 \times 10^{14} \mbox{cm}^{-3}\mbox{g}$ the nucleon density and $\rho$ the average density of the NS,
that for $W=U=7.9\times 10^{37}\mbox{erg/s}$ implies a change in the NS's temperature of $\Delta T\sim 10^3$K,
hence it is very hard to detect such changes of temperature in NS cooling observations. For the maximum AS mass
$M_{AS}=10^{-15}M_\odot$, the change in temperature will be  $\Delta T\sim 10^9$K, but the rate of collisions per
year is very small.

It may happen that the energy is not completely absorbed by the NS, but instead the axion-photon conversion
occurs out the NS's crust; perhaps on the NS's magnetosphere. If this scenario happens, some
photons may escape liberating the whole energy. But in this case, a particle
physics treatment for the oscillation may be required to compute the photon spectrum.

\section{Conclusions}
As we can see from Table \ref{Table2}, the energy dissipated by one collision of an AS with a
NS is many orders of magnitude bigger than the solar luminosity. But our results for the total dissipated energy per second in NS-AS collisions are around 10 orders of magnitude less than previous estimates \cite{Iwazaki:1999ub,Iwazaki:1999my}. As it can be seen from Fig \ref{Fig1}, this difference comes from our computation of the maximal AS mass obtained through resolution of EKG equations. The obtained maximal AS mass is much smaller than the corresponding parameter used in ref. \cite{Iwazaki:1999ub,Iwazaki:1999my}. It is important to notice that the maximal AS mass used in these references  are now  excluded by last observational results from Femtolensiong of Gamma Ray Bursts \cite{Barnacka:2012bm}.

Furthermore, improvement in femptolensing will help us to constrain the axion mass
by constraining the maximum mass of AS. Observation of a sudden release of energy in Neutron stars
may indicate the interaction of an AS with its strong magnetic field.
If some of those signatures are observed, they can be considered as indications that the
dark matter halo is made of an ensemble of axion stars. In our estimations, we have considered the
largest mass of the AS and a hypothetical electric
conductivity $\sigma$ \cite{Baiko:1995qg},  nevertheless,  we hope our results will encourage further
studies in the possibility of detecting the axion by its possible clustering.

\subsection*{Acknowledgements}
This work has been supported by CONACyT SNI-Mexico. The authors are also grateful to Conacyt (M\'exico) (CB-156618), DAIP project (Guanajuato University) and PIFI (Secretaria de Educacion Publica, M\'exico)
for financial support.
%%%%%%%%%%%%%%%%%%%%%%
%%%   REFERENCES   %%%
%%%%%%%%%%%%%%%%%%%%%%
\bibliography{referencias}
\bibliographystyle{apsrev}

%%%%%%%%%%%%%%%
%%%   END   %%%
%%%%%%%%%%%%%%%
\end{document}